\documentclass[12pt]{IEEEtran}
\usepackage{graphicx, amssymb}
\usepackage{graphicx}
\makeatletter
\def\ifundefined{\@ifundefined}
\makeatother

\begin{document}

\title{Impact of Channel Estimation Errors on Multiuser Detection via the Replica Method}

\author{Husheng Li and H. Vincent Poor\\\textit{Invited Paper}\thanks{Department of Electrical Engineering, Princeton University, Princeton, NJ 08544, USA ~(email:
\{hushengl, poor\}@princeton.edu). This research was supported in
part by the Office of Naval Research under Grant N00014-03-1-0102
and in part by the New Jersey Center for Wireless
Telecommunications. This paper was presented in part at the
  2004 IEEE Global Telecommunications Conference, Dallas, TX,
November 29 – December 3, 2004.d}}

\ifundefined{IEEEtransversionmajor}{%

   %
   \newlength{\IEEEilabelindent}
   \newlength{\IEEEilabelindentA}
   \newlength{\IEEEilabelindentB}
   \newlength{\IEEEelabelindent}
   \newlength{\IEEEdlabelindent}
   \newlength{\labelindent}
   \newlength{\IEEEiednormlabelsep}
   \newlength{\IEEEiedmathlabelsep}
   \newlength{\IEEEiedtopsep}

   \providecommand{\IEEElabelindentfactori}{1.0}
   \providecommand{\IEEElabelindentfactorii}{0.75}
   \providecommand{\IEEElabelindentfactoriii}{0.0}
   \providecommand{\IEEElabelindentfactoriv}{0.0}
   \providecommand{\IEEElabelindentfactorv}{0.0}
   \providecommand{\IEEElabelindentfactorvi}{0.0}
   \providecommand{\labelindentfactor}{1.0}

   \providecommand{\iedlistdecl}{\relax}
   \providecommand{\calcleftmargin}[1]{
                   \setlength{\leftmargin}{#1}
                   \addtolength{\leftmargin}{\labelwidth}
                   \addtolength{\leftmargin}{\labelsep}}
   \providecommand{\setlabelwidth}[1]{
                   \settowidth{\labelwidth}{#1}}
   \providecommand{\usemathlabelsep}{\relax}
   \providecommand{\iedlabeljustifyl}{\relax}
   \providecommand{\iedlabeljustifyc}{\relax}
   \providecommand{\iedlabeljustifyr}{\relax}

   \newif\ifnocalcleftmargin
   \nocalcleftmarginfalse

   \newif\ifnolabelindentfactor
   \nolabelindentfactorfalse

   \newif\ifcenterfigcaptions
   \centerfigcaptionsfalse

   \let\OLDitemize\itemize
   \let\OLDenumerate\enumerate
   \let\OLDdescription\description

   \renewcommand{\itemize}[1][\relax]{\OLDitemize}
   \renewcommand{\enumerate}[1][\relax]{\OLDenumerate}
   \renewcommand{\description}[1][\relax]{\OLDdescription}

   \providecommand{\pubid}[1]{\relax}
   \providecommand{\pubidadjcol}{\relax}
   \providecommand{\specialpapernotice}[1]{\relax}
   \providecommand{\overrideIEEEmargins}{\relax}

   \let\CMPARstart\PARstart

   \let\OLDappendix\appendix
   \renewcommand{\appendix}[1][\relax]{\OLDappendix}

   \newif\ifuseRomanappendices
   \useRomanappendicestrue

   \let\OLDbiography\biography
   \let\OLDendbiography\endbiography
   \renewcommand{\biography}[2][\relax]{\OLDbiography{#2}}
   \renewcommand{\endbiography}{\OLDendbiography}
   \markboth{A Test for IEEEtran.cls--- {\tiny \bfseries
   [Running Older Class]}}{Shell: A Test for IEEEtran.cls}}{

   \markboth{}%
   {Shell: A Test for IEEEtran.cls}}


%
%

\maketitle

\begin{abstract}
For practical wireless DS-CDMA systems, channel estimation is
imperfect due to noise and interference. In this paper, the impact
of channel estimation errors on multiuser detection (MUD) is
analyzed under the framework of the replica method. System
performance is obtained in the large system limit for optimal MUD,
linear MUD and turbo MUD, and is validated by numerical results for
finite systems.
\end{abstract}

\section{Introduction}
Multiuser detection (MUD) ~\cite{Verdu1998} can be used to
mitigate multiple access interference (MAI) in direct-sequence
code division multiple access (DS-CDMA) systems, thereby
substantially improving the system performance compared with the
conventional matched filter (MF) reception. The maximum likelihood
(ML) based optimal MUD, introduced in ~\cite{Verdu1984}, is
exponentially complex in the number of users, thus being difficult
to implement in practical systems. Consequently, various
suboptimal MUD algorithms have been proposed to effect a tradeoff
between performance and computational cost. For example, linear
processing can be applied, based on zero-forcing or minimum mean
square error (MMSE) criteria, thus resulting in the decorrelator
~\cite{Verdu1998} and the MMSE detector ~\cite{Lupas1989}. For
non-linear processing, a well known approach is decision feedback
based interference cancellation (IC) ~\cite{Verdu1998}, which can
be implemented in a parallel fashion (PIC) or successive fashion
(SIC). It should be noted that the above algorithms are suitable
for systems without channel codes. For channel coded CDMA systems,
the turbo principle can be introduced to improve the performance
iteratively using the decision feedback from channel decoders,
resulting in turbo MUD ~\cite{wang1999}, which can also be
simplified using PIC ~\cite{Alex2000}. The decisions of channel
decoders can also be fed back in the fashion of SIC, and it has
been shown that SIC combined with MMSE MUD achieves the sum
channel capacity ~\cite{Verdu1999}.

It is difficult to obtain explicit expressions for the performance
of most MUD algorithms in finite systems (Here, `finite' means
that the number of users and spreading gain are finite). In recent
years, asymptotic analysis has been applied to obtain the
performance of such systems in the large system limit, which means
that the system size tends to infinity while keeping the system
load a constant. The explicit expressions obtained from asymptotic
analysis can provide more insight than simulation results and can
be used as approximations for finite systems. The theory of large
random matrices ~\cite{Silverstein}~\cite{Voiculescu1992} has been
applied to the asymptotic analysis of MMSE MUD, resulting in the
Tse-Hanly equation ~\cite{Tse1999}, which quantifies implicitly
multiuser efficiency. However, this method is valid for only
linear MUD and cannot be used for the analysis of non-linear
algorithms. For ML optimal MUD, the performance is determined by
the sum of many exponential terms, which is difficult to tackle
with matrices. Recently, attention has been payed to the analogy
between optimal MUD and free energy in statistical mechanics
~\cite{Nishimori2001}, which has motivated researchers to apply
mathematical tools developed in statistical mechanics to the
analysis of MUD. In ~\cite{Tanaka2002}~\cite{GuoCISS2003}, the
replica method, which was developed in the context of spin glasses
theory, has been applied as a unified framework to both optimal
and linear MUD, resulting in explicit asymptotic expressions for
the corresponding bit error rates and spectral efficiencies. These
results have been extended to turbo MUD in ~\cite{Caire2003}. It
should be noted that the replica method is based on some
assumptions which still require rigorous mathematical proof.
However, the corresponding conclusions match simulation results
and some known theoretical conclusions well.

In practical wireless communication systems, the transmitted
signals experience fading. In the above MUD algorithms, the
channel state information (CSI) is assumed to be known to the
receiver. However, this is not a reasonable assumption since
channel estimation is imperfect due to the existence of noise and
interference. Therefore, it is of interest to analyze the
performance of MUD with imperfect channel estimates. For linear
MUD, the impact of channel estimation error on detection has been
studied in \cite{Evans2000}, \cite{ZhengyuanXu2004} and
\cite{LiHuWCNC2004} using the theory of large random matrices. In
this paper, we will apply the replica method to analyze the
corresponding impact on optimal MUD, and then extend the results
to linear or turbo MUD, under some assumptions on the channel
estimation error. The results can be used to determine the number
of training symbols needed for channel estimation.

The remainder of this paper is organized as follows. The signal
model is explained in Section II and the replica method is briefly
introduced in Section III. Optimal MUD with imperfect channel
estimation is discussed in Section IV and the results are extended
to linear and turbo MUD in Section V. Simulation results and
conclusions are given in Sections VI and VII, respectively.

\section{Signal Model}
\subsection{Signal Model}
We consider a synchronous uplink DS-CDMA system, which operates
over a frequency selective fading channel of order $P$ (i.e, $P$
is the delay spread in chip intervals). Let $K$ denote the number
of active users, $N$ the spreading gain and
$\beta\triangleq\frac{K}{N}$ the system load. In this paper, our
analysis is based on the large system limit, where
$K,N,P\rightarrow\infty$ while keeping $\frac{K}{N}$ and
$\frac{P}{N}$ constant.

We model the frequency selective fading channels as discrete
finite-impulse-response (FIR) filters. For simplicity, we assume
that the channel coefficients are real. The $z$-transform of the
channel response of user $k$ is given by
\begin{eqnarray}
h_k(z)=\sum_{p=0}^{P-1}g_k(p)z^p,
\end{eqnarray}
where $\{g_k(p)\}_{p=0,...,P-1}$ are the corresponding independent
and identically distributed (i.i.d.) (with respect to both $k$ and
$p$) channel coefficients having variance $\frac{1}{P}$. For
simplicity, we consider only the case in which $\frac{P}{N}\ll 1$.
Thus we can ignore the intersymbol interference (ISI) and deal
with only the portion uncontaminated by ISI.

The chip matched filter output at the $l$-th chip period in a
fixed symbol period can be written as
\begin{eqnarray}
r(l)=\frac{1}{\sqrt{N}}\sum_{k=1}^Kb_kh_k(l)+n(l), \qquad
l=P,P+1,...,N,
\end{eqnarray}
where $b_k$ denotes the binary phase shift keying (BPSK) modulated
channel symbol of user $k$ with normalized power 1, $\{n(l)\}$ is
additive white Gaussian noise (AWGN), which satisfies
$E\{|n(l)|^2\}=\sigma_n^2$ \footnote{Note that $\sigma_n^2$ is the
noise variance, normalized to represent the inverse
signal-to-noise ratio.} and $\{h_k(l)\}$ is the convolution of the
spreading codes and channel coefficients:
\begin{eqnarray}
h_k(l)=s_k(l)\star g_k(l),
\end{eqnarray}
where $s_k(l)$ is the $l$-th chip of the original spreading codes
of user $k$, which is i.i.d. with respect to both $k$ and $l$ and
takes values $1$ and $-1$ equiprobably. We call the $(N+P-1)\times
1$ vector $\textbf{h}_k=(h_k(0),...,h_k(N+P-2))^T$ the
\textit{equivalent spreading codes} of user $k$
\footnote{Superscript $T$ denotes transposition and superscript
$H$ denotes conjugate transposition.}. Due to the assumption that
$\frac{P}{N}\ll 1$, we can approximate $N-P+1$ by $N$ for
notational simplicity.
 Then the received signal in the fixed symbol period can be written in
a vector form:
\begin{eqnarray}
\textbf{r}=\frac{1}{\sqrt{N}}H\textbf{b}+\textbf{n},
\end{eqnarray}
where $\textbf{r}=\left(r(P),...,r(N)\right)^T$,
$H=\left(\textbf{h}_1,...,\textbf{h}_K\right)$ and
$\textbf{b}=\left(b_1,...,b_K\right)^T$.
It is easy to show that
$\frac{1}{N}\left\|\textbf{h}_k\right\|^2\rightarrow 1$, as
$P\rightarrow\infty$. Thus, we can ignore the performance loss
incurred by the fluctuations of received power in the fading
channels and consider only the impact of channel estimation error.

\subsection{Channel Estimation Error}
In practical wireless communication systems, the channel
coefficients $\left\{g_k(l)\right\}$ are unknown to the receiver,
and the corresponding channel estimates
$\left\{\hat{g}_k(l)\right\}$ are imprecise due to the existence
of noise and interference. We assume that training symbol based
channel estimation ~\cite{LiHuCISS2003} is applied to provide the
channel estimates. On denoting the channel estimation error by
$\delta g_k(l)\triangleq g_k(l)-\hat{g}_k(l)$, $\left\{\delta
g_k(l)\right\}$ are jointly Gaussian distributed and mutually
independent for sufficiently large numbers of training symbols
~\cite{LiHuCISS2003}. Therefore, it is reasonable to assume that
$\left\{\delta g_k(l)\right\}$ is independent for different values
of $k$ and $l$. In this paper, we consider only the following two
types of channel estimations.
\begin{itemize}
\item ML channel estimation. It is well known that the ML
estimation is asymptotically unbiased under some regulation
conditions. Thus, we can assume that the estimation error $\delta
g_k(l)$ has zero expectation conditioned on $g_k(l)$, and is
therefore correlated with $\hat{g}_k(l)$.

\item MMSE channel estimation. An important property of the MMSE
estimate, namely the conditional expectation
$E\left\{g_k(l)|Y\right\}$, where $Y$ is the observation, is that
the estimation error $\delta g_k(l)$ is uncorrelated with
$\hat{g}_k(l)$, and thus is biased.
\end{itemize}

We assume that the receiver uses the imperfect channel estimates
to construct the corresponding equivalent spreading codes, namely
$\hat{\textbf{h}}_k$. Thus, the error of the $i$-th chip of
$\hat{\textbf{h}}_k$ is given by
\begin{eqnarray}
\delta h_k(i)&\triangleq&h_k(i)-\hat{h}_k(i) \nonumber\\
             &=&\sum_{l=0}^{P-1} s_k(i-l)\delta g_k(l),
\end{eqnarray}
from which it follows that the variance of $\delta h_k(i)$ is
given by $\Delta_h^2=P\mbox{Var}\{\delta g_k(l)\}$.

Fixing $\left\{\delta g_k(l)\right\}$ and considering
$\left\{\delta s_k(l)\right\}$ as random variables, it is easy to
show that $\delta h_k(l)$ is asymptotically Gaussian as
$P\rightarrow \infty$ by applying the central limit theorem to
(5). Due to the assumption that $\frac{P}{N}\ll 1$, for any $l$,
$\delta h_k(l)$ is independent of most $\left\{\delta
h_k(m)\right\}_{m\neq l}$ since for any $|l-m|>P$, $\delta h_k(l)$
and $\delta h_k(m)$ are mutually independent. Thus, it is
reasonable to assume that the elements in $\delta \textbf{h}_k$
are Gaussian and mutually independent, which substantially
simplifies the analysis and will be validated with simulation
results in Section VI. Similarly, we can assume that the elements
of $\textbf{h}_k$ are mutually independent as well.

\section{Brief Review of Replica Method}
In this section, we give a brief introduction to the replica
method, on which the asymptotic analysis in this paper is based.
The details can be found in ~\cite{GuoCISS2002},
~\cite{GuoCISS2003}, ~\cite{Nishimori2001} and ~\cite{Tanaka2002}.

On assuming $P(b_k=1)=P(b_k=-1)$, we consider the following ratio
\begin{eqnarray}
\frac{P(b_k=1|\textbf{r})}{P(b_k=-1|\textbf{r})}=\frac{\sum_{\left\{\textbf{b}|b_k=1\right\}}\exp\left(-\frac{1}{2\sigma^2}\left\|\textbf{r}-\frac{1}{\sqrt{N}}H\textbf{b}\right\|^2\right)}{\sum_{\left\{\textbf{b}|b_k=-1\right\}}\exp\left(-\frac{1}{2\sigma^2}\left\|\textbf{r}-\frac{1}{\sqrt{N}}H\textbf{b}\right\|^2\right)},
\end{eqnarray}
where $\sigma^2$ is a control parameter. Various MUD algorithms
can be obtained using this ratio. In particular, we can obtain
individually optimal (IO), or maximum {\it a posteriori}
probability (MAP), MUD ($\sigma^2=\sigma_n^2$), jointly optimal
(JO), or ML, MUD ($\sigma^2=0$) and the MF ($\sigma^2=\infty$).

The key point of the replica method is the computation of the free
energy, which is given by
\begin{eqnarray}
\mathcal{F}_K(\textbf{r},H)&\triangleq&K^{-1}\log Z(\textbf{r},
H)\nonumber\\
                           &=&\lim_{K\rightarrow\infty}\int_{\mathbb{R}^N}
                           \overline{P(\textbf{r}|H)\log
                           Z(\textbf{r},H)}d\textbf{r},
\end{eqnarray}
where
$$
Z(\textbf{r},H)\triangleq
\sum_{\left\{\textbf{b}\right\}}P(\textbf{b})\exp\left(-\frac{1}{2\sigma^2}\left\|\textbf{r}-\frac{1}{\sqrt{N}}H\textbf{b}\right\|^2\right),
$$
and the overbar denotes the average over the randomness of the
equivalent spreading codes. It should be noted that the second
equation is based on the self-averaging assumption
~\cite{Tanaka2002}.

To evaluate the free energy, we can use the replica method, by
which we have
\begin{eqnarray}
\mathcal{F}_K(\textbf{r},H)=\lim_{K\rightarrow
\infty}\left(\lim_{n_r\rightarrow 0}\frac{\log\Xi_{n_r}}{K}\right),
\end{eqnarray}
where
\begin{eqnarray}
\Xi_{n_r}&=&\int_{\textbf{b}_0,...,\textbf{b}_{n_r}}\prod_{a=0}^{n_r}P(\textbf{b}_a)\nonumber\\
&\times&\left\{
\frac{1}{\sqrt{2\pi\sigma_n^2}}\int_{\mathbb{R}}\overline{\exp\left[-\frac{1}{2\sigma_n^2}\left(r-\frac{1}{\sqrt{N}}\sum_{k=1}^Kh_kb_{0k}\right)^2\right]\prod_{a=1}^{n_r}\exp\left[-\frac{1}{2\sigma^2}\left(r-\frac{1}{\sqrt{N}}\sum_{k=1}^Kh_kb_{ak}\right)^2\right]}
dr\right\}^N,\nonumber
\end{eqnarray}
where $\textbf{b}_0$ is the same as the $\textbf{b}$ in (4).
However, it is difficult to find an exact physical meaning for
$\left\{\textbf{b}_a\right\}_{a=1,...,n_r}$. We can roughly
consider $\textbf{b}_a$ to be the $a$-th estimates of the received
binary symbols $\textbf{b}$.

An assumption, which still lacks rigorous mathematical proof, is
proposed in ~\cite{Tanaka2002}, which states that $\Xi_{n_r}$
around $n_r=0$ can be evaluated by directly using the expression
of $\Xi_{n_r}$ obtained for positive integers $n_r$. With this
assumption, we can regard $n_r$ as an integer when evaluating
$\Xi_{n_r}$, and $\{\mathbf{x}_a\}$ as $n_r$ replicas of
$\mathbf{x}$.

To exploit the asymptotic normality of $\frac{1}{\sqrt{N}}\sum_{k=1}^K h_k
b_{ak}$, $a=0,...,n_r$, we define variables $\{v_a\}_{a=0,...,n_r}$ as
\begin{eqnarray}
\left\{\begin{array}{ll} v_0=\frac{1}{\sqrt{K}}\sum_{k=1}^K h_k
b_{0k},\\ v_a=\frac{1}{\sqrt{K}}\sum_{k=1}^K h_k b_{ak},\qquad
a=1,...,n_r. \end{array}\right.
\end{eqnarray}

The cross-correlations of $\left\{v_a\right\}$ are denoted by
parameters $\left\{Q_{ab}\right\}$, where $Q_{ab}\triangleq
\overline{v_av_b}$. With these definitions, we can obtain
\begin{eqnarray}
\Xi_{n_r}=\int_{\mathbb{R}}
\exp\left(K\beta^{-1}\mathcal{G}\left\{Q\right\}\right)\mu_K\left\{Q\right\}\prod_{a<b}dQ_{ab},
\end{eqnarray}
where\footnote{$\delta(x)$ is the Dirac delta function.}
$$
\mu_K\{Q\}=\sum_{\textbf{b}_0,...,\textbf{b}_{n_r}}\prod_{a=0}^{n_r}P(\textbf{b}_a)\prod_{a<b}\delta(\textbf{b}_a^H
\textbf{b}_b-KQ_{ab}),
$$
and
\begin{eqnarray}
\exp\left(\mathcal{G}\{Q\}\right)=\frac{1}{2\pi
\sigma_n^2}\int_{\mathbb{R}}
\overline{\exp\left[-\frac{\beta}{2\sigma_n^2}\left(\frac{r}{\sqrt{\beta}}-v_0\{Q\}\right)^2\right]\times
\prod_{a=1}^{n_r}\exp\left[-\frac{\beta}{2\sigma^2}\left(\frac{r}{\sqrt{\beta}}-v_a\{Q\}\right)^2\right]}dr+O(K^{-1}).\nonumber
\end{eqnarray}

By applying Varadhan's large deviations theorem
~\cite{Hollander2000}, $\Xi_{n_r}$ converges to the following
expression as $K\rightarrow\infty$:
\begin{eqnarray}
\lim_{K\rightarrow\infty}K^{-1}\log\Xi_{n_r}=\mbox{sup}_{\{Q\}}\left(\beta^{-1}\mathcal{G}\{Q\}-\mathcal{I}\{Q\}\right),
\end{eqnarray}
where $\mathcal{I}\{Q\}$ is the rate function of $\mu_K\{Q\}$,
which is based on an optimization over a set of
parameters $\{\tilde{Q}_{ab}\}_{a<b}$.

Thus, the evaluation of the free energy
$\mathcal{F}_K(\textbf{r},H)$ depends on the optimization of (11)
over the parameters $\{Q_{ab}\}$ and $\{\tilde{Q}_{ab}\}$, which
is computationally prohibitive. This problem is tackled by the
assumption of \textit{replica symmetry}; that is, $Q_{0a}=m$,
$\tilde{Q}_{0a}=E$, $\forall a\neq 0$ and $Q_{ab}=q$,
$\tilde{Q}_{ab}=F$, $\forall a<b,a\neq 0$. Then the optimization
of (11) is performed on the parameter set $\{m,q,E,F\}$. The
optimal $\{m,q,E,F\}$ are given by solving the following implicit
expressions:
\begin{eqnarray}
\left\{
\begin{array}{llll}
m=\int_\mathbb{R} \tanh\left(\sqrt{F}z+E\right)Dz\\
q=\int_\mathbb{R} \tanh^2\left(\sqrt{F}z+E\right)Dz\\
E=\frac{\beta^{-1}B}{1+B(1-q)}\\
F=\frac{\beta^{-1}B^2\left(B_0^{-1}+1-2m+q\right)}{(1+B(1-q))^2}
\end{array}
\right.,
\end{eqnarray}
where $Dz=\frac{1}{\sqrt{2\pi}}e^{-\frac{z^2}{2}}dz$,
$B_0=\frac{\beta}{\sigma_n^2}$ and $B=\frac{\beta}{\sigma^2}$. Then, the performance of MUD can be
derived from the free energy, which is determined by ${m,q,E,F}$. It
is shown in ~\cite{Tanaka2002} that the bit error rate of MUD is given by
\begin{eqnarray}
P_e=Q\left(\frac{{E}}{\sqrt{F}}\right),
\end{eqnarray}
where $Q(z)=\int_z^\infty Dt$ is the complementary Gaussian
cumulative distribution function. Thus the multiple access system
is equivalent to a single-user system operating over an AWGN
channel with an equivalent signal-to-noise ratio (SNR)
$\frac{E^2}{F}$. The parameters $m$ and $q$ are the first and
second moments, respectively, of the soft output,
$\hat{b}_k=P(b_k=1)-P(b_k=-1)$. When $B=B_0$
($\sigma^2=\sigma_n^2$), it is easy to check that $m=q$ and $E=F$
using (12).

\section{Optimal MUD}
In this section, we discuss two types of receivers distinguished
by whether or not the receiver considers the distribution of the
channel estimation error. We denote the case of
\underline{d}irectly using the channel estimates for MUD by a
prefix D, and the case of considering the distribution of the
channel estimation error to \underline{c}ompensate the
corresponding impact by a prefix C.

\subsection{D-optimal MUD}
In this subsection, we discuss the D-optimal MUD, where the
receiver applies the channel estimates directly to MUD and does not consider
the distribution of the channel estimation error. When the
equivalent spreading codes contain errors incurred by the channel
estimation error, the corresponding free energy is given by
\begin{eqnarray}
\mathcal{F}_K(\textbf{r},\hat{H})=K^{-1}\log
Z(\textbf{r},\hat{H}),
\end{eqnarray}
where $\hat{H}$ is the estimation of channel coefficients $H$ and
$$
Z(\textbf{r},\hat{H})\triangleq
\sum_{\left\{\textbf{b}\right\}}P(\textbf{b})\exp\left(-\frac{1}{2\sigma^2}\left\|\textbf{r}-\frac{1}{\sqrt{N}}\hat{H}\textbf{b}\right\|^2\right).
$$
We assume that the self-averaging assumption is also valid for
$\delta H\triangleq H-\hat{H}$, and thus (7) still holds with the
corresponding $\Xi_n$ given by
\begin{eqnarray}
\Xi_n&=&\int_{\textbf{b}_0,...,\textbf{b}_{n_r}}\prod_{a=0}^{n_r}P(\textbf{b}_a)\nonumber\\
&\times&\left\{
\frac{1}{\sqrt{2\pi\sigma_n^2}}\int_{\mathbb{R}}\overline{\exp\left[-\frac{1}{2\sigma_n^2}\left(r-\frac{1}{\sqrt{N}}\sum_{k=1}^Kh_kb_{0k}\right)^2\right]\prod_{a=1}^{n_r}\exp\left[-\frac{1}{2\sigma^2}\left(r-\frac{1}{\sqrt{N}}\sum_{k=1}^K\hat{h}_kb_{ak}\right)^2\right]}
dr\right\}^N.\nonumber
\end{eqnarray}

We can apply the same methodology as in Section III to the evaluation of the free
energy with imperfect channel estimation.
The only difference is that we need to take into account the
distribution of the channel estimation error. In a way similar to (9),
we define
$$
v_a=\frac{1}{\sqrt{K}}\sum_{k=1}^K \hat{h}_k b_{ak},\qquad
a=1,...,n_r.
$$

For ML channel estimation, $\delta h_k$ is uncorrelated with
$h_k$, thus resulting in $E\left\{h_k\hat{h}_k\right\}=1$ and
$E\left\{\hat{h}_k\hat{h}_k\right\}=1+\Delta_h^2$. Then we have
\begin{eqnarray}
\left\{\begin{array}{ll}
\overline{v_0v_a}=\frac{1}{K}\sum_{k=1}^Kb_{0k}b_{ak},
\qquad &\forall a>0, \\
\overline{v_av_b}=\frac{1+\Delta_h^2}{K}\sum_{k=1}^Kb_{ak}b_{bk},
\qquad &\forall a,b>0.\end{array}\right.
\end{eqnarray}

For MMSE channel estimation, $\delta h_k$ is uncorrelated with
$\hat{h}_k$, thus resulting in $E\{h_k
\hat{h}_k\}=E\{\hat{h}_k^2\}=1-\Delta_h^2$. Then we have
\begin{eqnarray}
\left\{\begin{array}{ll}
\overline{v_0v_a}=\frac{1-\Delta_h^2}{K}\sum_{k=1}^Kb_{0k}b_{ak},
\qquad &\forall a>0, \\
\overline{v_av_b}=\frac{1-\Delta_h^2}{K}\sum_{k=1}^Kb_{ak}b_{bk},
\qquad &\forall a,b>0.\end{array}\right.
\end{eqnarray}

Thus, the free energy with imprecise channel estimation still
depends on the same parameter set $\{m,q,E,F\}$ as in Section III.
An important observation is that the existence of $\{\delta h_k\}$
affects only the term $\mathcal{G}\{Q\}$ in (10), and $\mu_K\{Q\}$
remains unchanged, which implies that the expressions for $m$ and
$q$ are identical to those in (12). Hence, we can focus on only
the computation of $\mathcal{G}\{Q\}$. By supposing that the
assumption of replica symmetry is still valid, the asymptotically
Gaussian random variables $v_0$ and $v_a$ can be constructed using
expressions similar to those in ~\cite{Tanaka2002}. For ML channel
estimation, we have
\begin{eqnarray}
\left\{\begin{array}{ll}
v_0=u\sqrt{1-\frac{m^2}{(1+\Delta_h^2)q}}-t\frac{m}{\sqrt{(1+\Delta_h^2)q}},\\
v_a=\sqrt{1+\Delta_h^2}\left(z_a\sqrt{1-q}-t\sqrt{q}\right),\qquad a=1,...,n_r,\\
\end{array}\right.
\end{eqnarray}
where $u$, $t$ and $\left\{z_a\right\}$ are mutually independent Gaussian
random variables with zero mean and unit variance.

With the same definitions of $u$, $t$ and $\left\{z_a\right\}$,
for MMSE channel estimation, we have
\begin{eqnarray}
\left\{\begin{array}{ll}
v_0=u\sqrt{1-\frac{(1-\Delta_h^2)m^2}{q}}-t\frac{m\sqrt{1-\Delta_h^2}}{\sqrt{q}},\\
v_a=\sqrt{1-\Delta_h^2}\left(z_a\sqrt{1-q}-t\sqrt{q}\right),\qquad a=1,...,n_r.\\
\end{array}\right.
\end{eqnarray}

Substituting the above expressions into (10), we can obtain the
following conclusions using some calculus similar to that of
~\cite{Tanaka2002}. For ML channel estimation, the free energy is
given by
\begin{eqnarray}
\mathcal{F}_K\left(\textbf{r},\hat{H}\right)&=&\int_{\mathbb{R}}
\log\left(\cosh\left(\sqrt{F}z+E\right)\right)Dz-Em-\frac{F(1-q)}{2}\nonumber\\
&-&\frac{1}{2\beta}\left(\log\left(1+\left(1+\Delta_h^2\right)(1-q)B\right)+\frac{B\left(B_0^{-1}+1-2m+(1+\Delta_h^2)q\right)}{1+B(1-q)(1+\Delta_h^2)}\right).
\end{eqnarray}

The corresponding $E$ and $F$ are given by
\begin{eqnarray}
\left\{\begin{array}{ll}
E=\frac{\beta^{-1}B}{1+B(1-q)(1+\Delta_h^2)},\\
F=\frac{(1+\Delta_h^2)\beta^{-1}B^2\left(B_0^{-1}+1-2m+(1+\Delta_h^2)q\right)}{(1+B(1-q)(1+\Delta_h^2))^2}.
\end{array}\right.
\end{eqnarray}

For MMSE channel estimation, we can obtain
\begin{eqnarray}
\mathcal{F}_K\left(\textbf{r},\hat{H}\right)&=&\int_{\mathbb{R}}
\log\left(\cosh\left(\sqrt{F}z+E\right)\right)Dz-Em-\frac{F(1-q)}{2}\nonumber\\
&-&\frac{1}{2\beta}\left(\log\left(1+\left(1-\Delta_h^2\right)(1-q)B\right)+\frac{B\left(B_0^{-1}+1-(1-\Delta_h^2)(2m-q)\right)}{1+B(1-q)(1-\Delta_h^2)}\right),
\end{eqnarray}
and the corresponding $E$ and $F$ are given by
\begin{eqnarray}
\left\{\begin{array}{ll}
E=\frac{\beta^{-1}B(1-\Delta_h^2)}{1+B(1-q)(1-\Delta_h^2)},\\
F=\frac{\beta^{-1}B^2(1-\Delta_h^2)\left(B_0^{-1}+1-(1-\Delta_h^2)(2m-q)\right)}{(1+B(1-q)(1-\Delta_h^2))^2}.
\end{array}\right.
\end{eqnarray}

The corresponding output signal-to-interference-plus-noise-ratios
(SINRs) of the ML and MMSE channel estimation are given by the
following expressions, respectively.
\begin{eqnarray}
\mbox{SINR}_{ML}=
\frac{1}{\left(1+\Delta_h^2\right)}\frac{1}{\left(\sigma_n^2+\beta\left(1-2m+(1+\Delta_h^2)q\right)\right)},
\end{eqnarray}
and
\begin{eqnarray}
\mbox{SINR}_{MMSE}=
\frac{1-\Delta_h^2}{\left(\sigma_n^2+\beta\left(1-(1-\Delta_h^2)(2m-q)\right)\right)}.
\end{eqnarray}

Thus, we can summarize the impact of the channel estimation error
on the D-optimal MUD as follows:
\begin{itemize}
\item The factors $\frac{1}{1+\Delta_h^2}$ in (23) and
$1-\Delta_h^2$ in the numerator of (24) represent the impact of
the error of the desired user's equivalent spreading codes, which
is equivalent to increasing the noise level.

\item The imperfect channel estimation also increases the variance
of the residual MAI, which equals $\beta(1-2m+(1+\Delta_h^2)q)$
for ML channel estimation based systems and
$\beta(1-(1-\Delta_h^2)(2m-q))$ for MMSE channel estimation based
systems.

\item The equations that $m=q$ and $E=F$ are no longer valid when
$\sigma^2=\sigma_n^2$. Thus, there are no simple analytical
expressions for obtaining the multiuser efficiency in a similar
way to the Tse-Hanly equation ~\cite{Tse1999}.
\end{itemize}

\subsection{C-optimal MUD}
In this subsection, we consider the C-optimal MUD, where the
distribution of the channel estimation error is exploited to
compensate for the imperfection of channel estimation. For
simplicity, we consider only the IO MUD (C-IO MUD).

\subsubsection{ML Channel Estimation}
When deriving the expressions of C-IO MUD, we consider
a fixed chip period and drop the index of the chip period
for simplicity. The conditional probability
$P\left(\left\{h_k\right\}\bigg|\left\{\hat{h}_k\right\}\right)$ should
be taken into account to attain the optimal detection. Thus, the
\textit{a posteriori} probability of the received signal $r$ at
this chip period, conditioned on the channel estimates
$\left\{\hat{h}_k\right\}$ and the transmitted symbols
$\left\{b_k\right\}$, is given by
\begin{eqnarray}
P\left(r\bigg|\left\{\hat{h}_k\right\},\{b_k\}\right)\propto\int_{\mathbb{R}^K}
P\left(r\bigg|\left\{h_k\right\},\{b_k\}\right)P\left(\left\{h_k\right\}\bigg|\left\{\hat{h}_k\right\}\right)\prod_{k=1}^Kdh_k,
\end{eqnarray}
where
$$
P\left(\left\{h_k\right\}\bigg|\left\{\hat{h}_k\right\}\right)=\prod_{k=1}^KP\left(h_k\bigg|\hat{h}_k\right),
$$
and
$$
P\left(h_k|\hat{h}_k\right)\propto
\exp\left(-\frac{\left(h_k-\hat{h}_k\right)^2}{2\Delta_h^2}\right)\exp\left(-\frac{h_k^2}{2}\right).
$$
It should be noted that the above two expressions are based on the assumption of normality and
mutual independence of $\{\delta h_k\}$ in Section II.B. Then we have
\begin{eqnarray}
P\left(r\bigg|\left\{\hat{h}_k\right\},\{b_k\}\right)\propto
\int_{\mathbb{R}^K}\exp\left(-\frac{\left(r-\frac{1}{\sqrt{N}}\sum_{k=1}^Kh_kb_k\right)^2}{2\sigma_n^2}\right)\prod_{k=1}^K
p(h_k|\hat{h}_k)dh_k.
\end{eqnarray}

Let $r_1=r-\frac{1}{\sqrt{N}}\sum_{k=2}^Kh_kb_k$, then the
integral with respect to $h_1$ is given by
\begin{eqnarray}
&&\int_{\mathbb{R}}
\exp\left(-\frac{\left(r_1-\frac{1}{\sqrt{N}}h_1b_1\right)^2}{2\sigma_n^2}\right)\exp\left(-\frac{\left(h_1-\hat{h}_1\right)^2}{2\Delta_h^2}\right)\exp\left(-\frac{h_1^2}{2}\right)dh_1\nonumber\\
&\propto&\exp\left(-\frac{\left(r_1-\frac{b_1\hat{h}_1}{\sqrt{N}(1+\Delta_h^2)}\right)^2}{2\left(\sigma_n^2+\frac{\Delta_h^2}{(1+\Delta_h^2)N}\right)}\right),
\end{eqnarray}
where the factors common for different $\{b_k\}$ are ignored for
simplicity.

Applying the same procedure for $h_2$, ..., $h_K$, we obtain that
\begin{eqnarray}
P\left(r\bigg|\left\{\hat{h}_k\right\},\{b_k\}\right)\propto
\exp\left(-\frac{\left(r-\frac{1}{\sqrt{N}(1+\Delta_h^2)}\sum_{k=1}^Kb_k\hat{h}_k\right)^2}{2\left(\sigma_n^2+\frac{\beta\Delta_h^2}{1+\Delta_h^2}\right)}\right).
\end{eqnarray}

Thus the LR of IO MUD is given by
\begin{eqnarray}
\frac{P(b_k=1|\textbf{r})}{P(b_k=-1|\textbf{r})}=\frac{\sum_{\left\{\textbf{b}|b_k=1\right\}}\exp\left(-\frac{1}{2\sigma^2}\left\|\textbf{r}-\frac{1}{\sqrt{N}(1+\Delta_h^2)}\hat{H}\textbf{b}\right\|^2\right)}{\sum_{\left\{\textbf{b}|b_k=-1\right\}}\exp\left(-\frac{1}{2\sigma^2}\left\|\textbf{r}-\frac{1}{\sqrt{N}(1+\Delta_h^2)}\hat{H}\textbf{b}\right\|^2\right)},
\end{eqnarray}
where $\sigma^2=\sigma_n^2+\frac{\beta\Delta_h^2}{1+\Delta_h^2}$.
Therefore, the channel estimation error is compensated for merely
by changing the equivalent noise variance and scaling the channel
estimate with a factor of $\frac{1}{1+\Delta_h^2}$.

Similarly to the analysis in Section IV.A, we can define
\begin{eqnarray}
\left\{\begin{array}{ll}
v_0=u\sqrt{1-\frac{m^2}{(1+\Delta_h^2)q}}-t\frac{m}{\sqrt{(1+\Delta_h^2)q}},\\
v_a=\frac{1}{\sqrt{1+\Delta_h^2}}\left(z_a\sqrt{1-q}-t\sqrt{q}\right),\qquad a=1,...,n_r.\\
\end{array}\right.
\end{eqnarray}

Then we can obtain the free energy, which is given by
\begin{eqnarray}
\mathcal{F}_K\left(\textbf{r},\hat{H}\right)&=&\int_{\mathbb{R}}
\log\left(\cosh\left(\sqrt{F}z+E\right)\right)Dz-Em-\frac{F(1-q)}{2}\nonumber\\
&-&\frac{1}{2\beta}\left(\log\left(1+\frac{B(1-q)}{\left(1+\Delta_h^2\right)}\right)+\frac{B\left(\left(B_0^{-1}+1\right)(1+\Delta_h^2)-2m+q\right)}{1+\Delta_h^2+B(1-q)}\right),
\end{eqnarray}
where
$B=\frac{\beta}{\sigma_n^2+\frac{\beta\Delta_h^2}{1+\Delta_h^2}}$.
The corresponding $E$ and $F$ are given by
\begin{eqnarray}
\left\{\begin{array}{ll}
E=\frac{\beta^{-1}B_0}{1+\Delta_h^2+B_0(1+\Delta_h^2-q)},\\
F=\frac{\beta^{-1}B_0^2\left((B_0^{-1}+1)(1+\Delta_h^2)-2m+q\right)}{\left(1+\Delta_h^2+B_0(1+\Delta_h^2-q)\right)^2}.
\end{array}\right.
\end{eqnarray}

An interesting observation is that the equations $m=q$ and $E=F$
are recovered in this case. Also we can obtain the equivalent
SINR, which is given by
\begin{eqnarray}
\mbox{SINR}_{ML}=\frac{1}{\sigma_n^2(1+\Delta_h^2)+\beta\Delta_h^2+\beta(1-q)}.
\end{eqnarray}

The corresponding multiuser efficiency $\eta$ is given by solving the
following Tse-Hanly style equation:
\begin{eqnarray}\
\frac{1}{\eta}+\frac{\beta}{\sigma_n^2}\int_{\mathbb{R}}
\tanh^2\left(\sqrt{\frac{\eta}{\sigma_n^2}}z+\frac{\eta}{\sigma_n^2}\right)Dz=\left(1+\Delta_h^2\right)\left(1+\frac{\beta}{\sigma_n^2}\right).
\end{eqnarray}

From (33), we can see that the impact of channel estimation error
consists of three aspects, which are represented by the three
terms in the denominator of the expression (33). The term
$\sigma_n^2\left(1+\Delta_h^2\right)$ embodies the negative impact
of the channel estimation error on the user being detected, which
causes uncertainty in the equivalent spreading codes of this user
and is equivalent to scaling the noise by a factor of
$\left(1+\Delta_h^2\right)$. Besides implicitly affecting the
parameter $q$ in the third term, the channel estimation error of
the interfering users also results in the term of
$\beta\Delta_h^2$; an intuitive explanation for this is that,
since the output of IO MUD can be regarded as the output of an
interference canceller using the conditional mean estimates of all
other users ~\cite{GuoCISS2003}, the channel estimation error
causes imperfection in the reconstruction of the signals of the
other users and the variance of residual interference equals
$\beta\Delta_h^2$ when the decision feedback is free of errors.

\subsubsection{MMSE Channel Estimation}
For MMSE channel estimation, the channel estimation error $\delta
h_k$ is uncorrelated with the estimate $\hat{h}_k$. Thus, we have
\begin{eqnarray}
P\left(h_k\bigg|\hat{h}_k\right)&=&P\left(\delta h_k+\hat{h}_k\bigg|\hat{h}_k\right)\nonumber\\
            &\propto&\exp\left(-\frac{(h_k-\hat{h}_k)^2}{2\Delta_h^2}\right).
\end{eqnarray}
Applying the same procedure as ML channel estimation, we can
obtain the LR of IO MUD, which is given by
\begin{eqnarray}
\frac{P(b_k=1|\textbf{r})}{P(b_k=-1|\textbf{r})}=\frac{\sum_{\left\{\textbf{b}|b_k=1\right\}}\exp\left(-\frac{1}{2\sigma^2}\left\|\textbf{r}-\frac{1}{\sqrt{N}}\hat{H}\textbf{b}\right\|^2\right)}{\sum_{\left\{\textbf{b}|b_k=-1\right\}}\exp\left(-\frac{1}{2\sigma^2}\left\|\textbf{r}-\frac{1}{\sqrt{N}}\hat{H}\textbf{b}\right\|^2\right)},
\end{eqnarray}
where the control parameter, or equivalent noise power, $\sigma^2=\sigma_n^2+\beta\Delta_h^2$. Substituting
$B=\frac{\beta}{\sigma_n^2+\beta\Delta_h^2}$ into (22), we have
\begin{eqnarray}
\left\{\begin{array}{ll}
E=\frac{\beta^{-1}B_0(1-\Delta_h^2)}{1+B_0(1-(1-\Delta_h^2)q)}\\
F=\frac{\beta^{-1}B_0^2(1-\Delta_h^2)\left(B_0^{-1}-(2m-q)(1-\Delta_h^2)\right)}{\left(1+B_0(1-(1-\Delta_h^2)q)\right)^2}
\end{array}\right..
\end{eqnarray}
Similarly to the case of ML channel estimation, the equations
$m=q$ and $E=F$ are recovered as well. The equivalent output SINR
is given by
\begin{eqnarray}
\mbox{SINR}_{MMSE}=\frac{1-\Delta_h^2}{\sigma_n^2+\beta(1-(1-\Delta_h^2)q)},
\end{eqnarray}
and the corresponding multiuser efficiency is given by solving the
following equation:
\begin{eqnarray}
\frac{1}{\eta}+\frac{\beta}{\sigma_n^2}\int_{\mathbb{R}}\tanh^2\left(\sqrt{\frac{\eta}{\sigma_n^2}}z+\frac{\eta}{\sigma_n^2}\right)=\frac{1+\frac{\beta}{\sigma_n^2}}{1-\Delta_h^2}.
\end{eqnarray}

The intuition behind (38) is similar to that of ML channel
estimation. On comparing (34) and (39), an immediate conclusion is
that the C-IO MUD is more susceptible to the error incurred by
MMSE channel estimation than that incurred by ML channel
estimation, when $\Delta_h^2$ is identical for both estimators.

\section{Linear MUD and Turbo MUD}
We now turn to the consideration of linear and turbo multiuser
detection. For simplicity, we discuss only ML channel estimation
based systems in this section. MMSE channel estimation based
systems can be analyzed in a similar way.

\subsection{Linear MUD}
The analysis of linear MUD can be incorporated into the framework
of the replica method (for MMSE MUD, $\sigma^2=\sigma_n^2$; for
the decorrelator, $\sigma^2\rightarrow 0$) by merely regarding the
channel symbols as Gaussian distributed random variables. The
system performance is determined by the parameter set $\{m, q, p,
E, F, G\}$ and a group of saddle-point equations
~\cite{Tanaka2002}.

Particularly, when $\sigma^2=\sigma^2_n$ (MMSE MUD), the
parameters can be simplified to $\{q,E\}$, which satisfy
$q=\frac{E}{1+E}$ and $E=\frac{\beta^{-1}B_0}{1+B_0(1-q)}$. The
multiuser efficiency is determined by the Tse-Hanly equation
~\cite{Tse1999}.

\subsubsection{D-MMSE MUD}
Since the channel estimation error does not affect
$\mathcal{I}\{Q\}$, the parameters $m$, $q$ and $p$ are unchanged.
With the same manipulation on $\mathcal{G}\{Q\}$ as in
Section IV, we can obtain the parameters $E$, $F$ and $G$
as follows:
\begin{eqnarray}
\left\{\begin{array}{lll}
E=\frac{\beta^{-1}B}{1+B(p-q)(1+\Delta_h^2)},\\
F=\frac{(1+\Delta_h^2)\beta^{-1}B^2\left(B_0^{-1}+1-2m+(1+\Delta_h^2)q\right)}{(1+B(p-q)(1+\Delta_h^2))^2},\\
G=F-\left(1+\Delta_h^2\right)E.
\end{array}\right.
\end{eqnarray}

\subsubsection{C-MMSE MUD}
Similarly to Section IV, the MMSE detector considering of the
distribution of the channel estimation error is given by merely
scaling $\hat{H}$ with a factor of $\frac{1}{1+\Delta_h^2}$ and
changing $\sigma^2$ to $\sigma_n^2+\frac{\beta
\Delta_h^2}{1+\Delta_h^2}$. Then, we have $E=F$, $G=0$, $m=q$ and
$p=0$. The corresponding multiuser efficiency is given implicitly
by
\begin{eqnarray}
\left(1+\Delta_h^2+\frac{\beta\Delta_h^2}{\sigma_n^2}\right)\eta+\frac{\beta\eta}{\sigma_n^2+\eta}=1.
\end{eqnarray}

\subsection{Turbo MUD}
\subsubsection{Optimal turbo MUD}
For optimal turbo MUD ~\cite{wang1999}, since the channel
estimation error does not affect $\mathcal{I}\{Q\}$ when
evaluating the free energy, the impact of channel estimation error
is similar to the optimal MUD in Section IV, namely, the
corresponding saddle-point equations remain the same as in
~\cite{Caire2003} except that the parameters $E$ and $F$ are
changed in the same way as in (20) and (32).

\subsubsection{MMSE filter based PIC}
However, greater complications arise in the case of MMSE filter
based PIC ~\cite{wang1999}, where the MAI is cancelled with the
decision feedback from channel decoders and the residual MAI is
further suppressed with an MMSE filter. The corresponding MMSE
filter is constructed with the estimated equivalent spreading
codes $\left\{\hat{\textbf{h}}_k\right\}$ and the estimated power
of the residual interference. In an unconditional MMSE filter, the
power estimate is given by $\Delta_b^2\triangleq
E\left\{\left(b_k-\hat{b}_k\right)^2\right\}$, where $\hat{b}_k$
is the soft decision feedback; and in a conditional MMSE filter,
the power estimate is given by $1-\hat{b}_k^2$. However, this
power estimate for user $k$ is different from the true value
$\left|b_k-\hat{b}_k\right|^2$ since $b_k$ is unknown to the
receiver, thus making the filter unmatched for the MAI. Hence, the
analysis in ~\cite{Caire2003} may overestimate the system
performance since such power estimation errors are not considered
there. Thus we need to take into account the corresponding power
mismatch. For simplicity, we consider only unbiased power
estimation. Note that this scenario can be applied to general
cases where the received signal power is not perfectly estimated.

For the MMSE filter based PIC, the powers of the residual
interference are different for different users. Similarly to the
analysis of unequal-power systems in ~\cite{GuoCISS2002}, we can
divide the users into a finite number ($L$) of equal-power groups,
with power $\left\{P_l\right\}_{l=1,...,L}$, estimated power
$\left\{\hat{P}_l\right\}_{l=1,...,L}$ and the corresponding
proportion $\left\{\alpha_l\right\}_{l=1,...,L}$, and obtain the
results for any arbitrary user power distribution by letting
$L\rightarrow\infty$. Confining our discussion to unbiased MAI
power estimation, we normalize the MAI power such that
$\sum_{l=1}^L\alpha_l P_l=1$ and $\sum_{l=1}^L\alpha_l
\hat{P}_l=1$. The equivalent noise variance is given by
$\sigma^2=\frac{\sigma_n^2}{\Delta_b^2}$. Thus, the bit error rate
of MUD is given by $Q\left(\frac{E}{\sqrt{F\Delta_b^2}}\right)$
since the power of the desired user is unity.

Similarly to the previous analysis, we define
\begin{eqnarray}
v_0&=&\frac{1}{\sqrt{K}}\sum_{l=1}^L\sqrt{P_k}\sum_{k\in
C_l}h_kb_{0k},\nonumber\\
v_a&=&\frac{1}{\sqrt{K}}\sum_{l=1}^L\sqrt{\hat{P}_k}\sum_{k\in
C_l}\hat{h}_kb_{ak},\qquad a=1,...,n_r,\nonumber
\end{eqnarray}
where $C_l$ represents the set of users with power $P_l$. We can
see that the uneven and mismatched power distribution does not
affect the analysis of $\exp\left(\mathcal{G}\{Q\}\right)$, which incorporates the impact of channel estimation error.
However, the rate function $\mathcal{I}\{Q\}$ is changed to
\begin{eqnarray}
\mathcal{I}\{Q\}=\sup_{\{\tilde{Q}\}}\left(\sum_{a\leq
b}\tilde{Q}_{ab}Q_{ab}-\sum_{l=1}^L\alpha_l\log
M^{G}_{\{l\}}\{\tilde{Q}\}\right),
\end{eqnarray}
where
\begin{eqnarray}
M^{G}_{\{l\}}\{\tilde{Q}\}=\frac{1}{2}\int_{\mathbb{R}^{n_r}}\exp\left(\sqrt{P_l\hat{P}_l}Eb_0\sum_{a=1}^{n_r}b_a+\hat{P}_lF\sum_{a<b}b_ab_b+\frac{G\hat{P}_l}{2}\sum_{a=1}^{n_r}b_a^2\right)\prod_{a=1}^n
Db_a,
\end{eqnarray}
in which $\{b_a\}_{a=1,...,n_r}$ are Gaussian random variables.
Similarly to ~\cite{GuoCISS2002}, after some algebra, we can
obtain the free energy, which is given by

\begin{eqnarray}
\mathcal{F}_K\left(\textbf{r},\hat{H}\right)&=&\frac{1}{2}\sum_{l=1}^L\alpha_l\left(\log\left(1+(F-G)\hat{P}_l\right)-\frac{\hat{P}_lF+P_l\hat{P}_lE^2}{1+(F-G)\hat{P}_l}\right)+Em-\frac{1}{2}Fq+\frac{1}{2}Gp\nonumber\\
&-&\frac{1}{2\beta}\left(\log\left(1+\left(1+\Delta_h^2\right)(p-q)B\right)+\frac{B\left(B_0^{-1}+1-2m+(1+\Delta_h^2)q\right)}{1+B(p-q)(1+\Delta_h^2)}\right).
\end{eqnarray}

Letting $L\rightarrow\infty$, we can obtain that
\begin{eqnarray}
\left\{\begin{array}{lll}
m=E\left\{\frac{P\hat{P}E}{1+\hat{P}(F-G)}\right\}\\
q=E\left\{\frac{\hat{P}^2(PE^2+F)}{\left(1+\hat{P}(F-G)\right)^2}\right\}\\
p=E\left\{\frac{\hat{P}\left(\hat{P}PE^2+2\hat{P}F+1-\hat{P}G\right)}{\left(1+\hat{P}(F-G)\right)^2}\right\}
\end{array}\right.,
\end{eqnarray}
where the expectation is with respect to the joint distribution of
$P$ and $\hat{P}$.

For the unconditional MMSE filter, the expressions for $m$, $q$
and $p$ can be simplified to the following expressions, since
$\hat{P}=E\{P\}=\Delta_b^2$:
\begin{eqnarray}
\left\{\begin{array}{lll}
m=\frac{\left(\Delta_b^2\right)^2E}{1+\Delta_b^2(F-G)}\\
q=\frac{\left(\Delta_b^2\right)^2(\Delta_b^2E^2+F)}{\left(1+\Delta_b^2(F-G)\right)^2}\\
p=\frac{\Delta_b^2\left(\left(\Delta_b^2\right)^2E^2+2\Delta_b^2F+1-\Delta_b^2G\right)}{\left(1+\Delta_b^2(F-G)\right)^2}
\end{array}\right..
\end{eqnarray}

This implies the interesting conclusion that if the MMSE MUD based
receiver regards the received powers of different users as being
equal to the average received power, the multiuser efficiency will
be identical to that of the corresponding equal-power system. It
should be noted that the corresponding bit error rates are
different although the multiuser efficiencies are the same. Thus,
the analysis of the unconditional MMSE filter based PIC in
~\cite{Caire2003} yields correct results. It should be noted that,
for IO MUD with binary channel symbols, this conclusion does not
hold since the expressions for $m$, $q$ and $p$ are nonlinear
in$P$.

This conclusion can also be applied to frequency-flat fading channels. When
the received power is perfectly known, the multiuser efficiency of
MMSE MUD is given by
\begin{eqnarray}
\eta+E\left\{\frac{\beta P\eta}{\sigma_n^2+P\eta}\right\}=1,
\end{eqnarray}
where the random variable $P$ is the received power and the
expectation is with respect to the distribution of $P$. When the
receiver is unaware of the fading and uses equal-power MMSE MUD,
the multiuser efficiency of this power-mismatched MMSE MUD is
given by that of an equal-power system:
\begin{eqnarray}
\eta+\frac{\beta E\{P\}\eta}{\sigma_n^2+E\{P\}\eta}=1.
\end{eqnarray}
Comparing (47) and (48) and applying the fact that, for any
positive random variable $x$, $E\left\{\frac{x}{1+x}\right\}\leq
\frac{E\{x\}}{1+E\{x\}}$, we can see that this power mismatch
incurs a loss in multiuser efficiency.

\section{Simulation Results}
In this section, we provide simulation results to verify and
illustrate the analysis of the preceding sections.

Figure 1 shows the bit error rates versus the variance of the
channel estimation error for a D-IO MUD system with $K=10$,
$N=150$, $P=50$ and $\sigma_n^2=0.2$. In this figure,
`independent' represents the case of equivalent spreading codes
with mutually independent elements and `convolution' represents
the case in which the equivalent spreading codes are the
convolutions of binary spreading codes and channel gains. From
this figure, we can see that the assumption of independent
elements in the equivalent spreading codes appears to be valid and
the asymptotic results can predict the performance of finite
systems fairly well. This figure also shows that D-IO MUD is more
susceptible to the error of MMSE channel estimation than that of
ML estimation.

Figure 2 compares the bit error rates in D-IO and C-IO MUD systems
with $\beta=0.5$ and $\sigma_n^2=0.2$. For ML channel estimation,
the C-IO MUD achieves considerably better performance than the
D-IO MUD. For MMSE channel estimation, the two IO MUD schemes
attain almost the same performance.

Figure 3 shows the bit error rates for MMSE MUD systems with the
same configuration as in Fig. 3. Both the numerical simulations
(for both independent and convolution models of the equivalent
spreading codes) and asymptotic results are given for D-MMSE MUD,
and match fairly well. Note that C-MMSE MUD achieves marginally
better performance than D-MMSE MUD.

Figure 4 shows the bit error rates of MMSE filter based PIC
systems with the same configurations as in Fig.3. The decision
feedback is from the channel decoder of convolutional code
$(\mbox{23,33,37})_8$ when the input SINR is 3dB. In this figure,
the theoretical and simulation results for the unconditional MMSE
filter are represented with `mismatched' and `simulation',
respectively; the results with the assumption that the residual
interference power is known are represented by `optimal'. We can
observe that the optimal scheme, which assumes that the decision
feedback error is known, achieves only marginally better
performance.

For Rayleigh flat-fading channels, the multiuser efficiency,
obtained by numerical simulations, versus SNR is given in Fig. 5.
In this figure, `Equal Power' means the case of equal received
power. For the case of Rayleigh distributed received power, the
results of mismatched (regarding the received power as being
equal) MMSE MUD and optimal (the received powers are known) MMSE
MUD are represented by `Rayleigh-Mismatch' and `Rayleigh',
respectively. We can see that the numerical results verify our
conclusion about the power-mismatched MMSE MUD in Section V.B.
Also, the knowledge of received power provides marginal
improvement in multiuser efficiency.

In Fig. 6, we apply the results for C-MMSE MUD to obtain the
optimal proportion $\alpha$ of training symbols, versus different
coherence time $M$ (measured in symbol periods) and system load
$\beta$, to maximize the spectral efficiency given by
$(1-\alpha)\log(1+\eta\mbox{SNR})$, where $\mbox{SNR}=5$dB, $\eta$
is determined by (41) and $\Delta_h^2=\frac{\sigma_n^2}{\alpha
M}$. We can see that the required proportion of training data
increases with the system load and decreases with the coherence
time.

\section{Conclusions}
In this paper, we have discussed the impact of channel estimation
error on various types of MUD algorithms in DS-CDMA systems by
obtaining the asymptotic expressions of the system
performance in terms of the channel estimation error variance. The
analysis is unified under the framework of the replica method. The
following conclusions are of particular interest:
\begin{itemize}
\item The performance of MUD is more susceptible to MMSE channel
estimation errors than ML ones.

\item The MUD schemes that consider the distribution of channel
estimation errors can improve the system performance, considerably
for ML channel estimation errors and marginally for MMSE channel
estimation errors.

\item When the MMSE MUD treats different users as being
received with equal power, it attains the same multiuser efficiency as
the corresponding equal-power systems.
\end{itemize}

\begin{figure}
  \centering
  \includegraphics[scale=1]{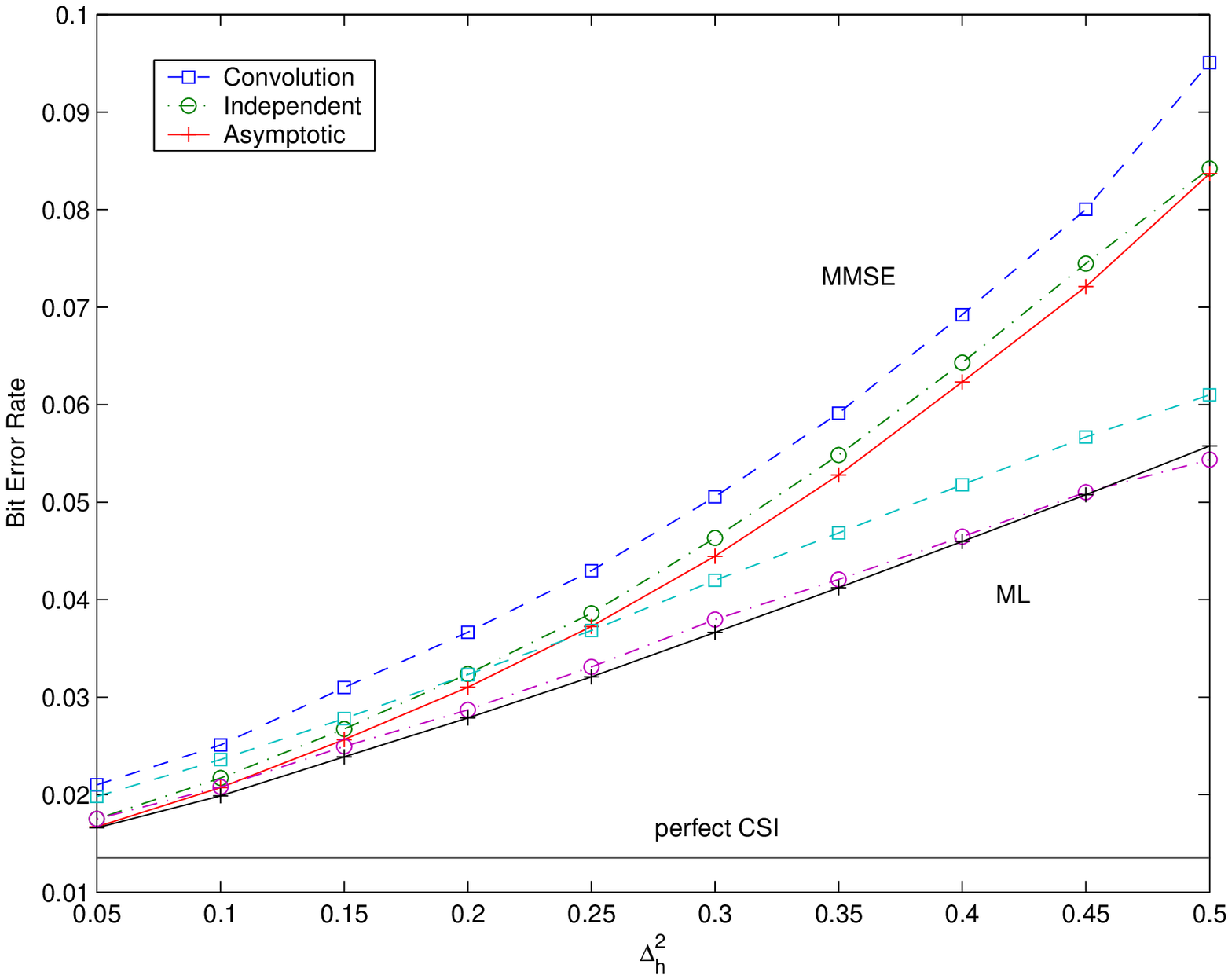}
  \caption{Bit error rate of D-IO MUD as a function of channel estimation error variance}\label{}
\end{figure}

\begin{figure}
  \centering
  \includegraphics[scale=1]{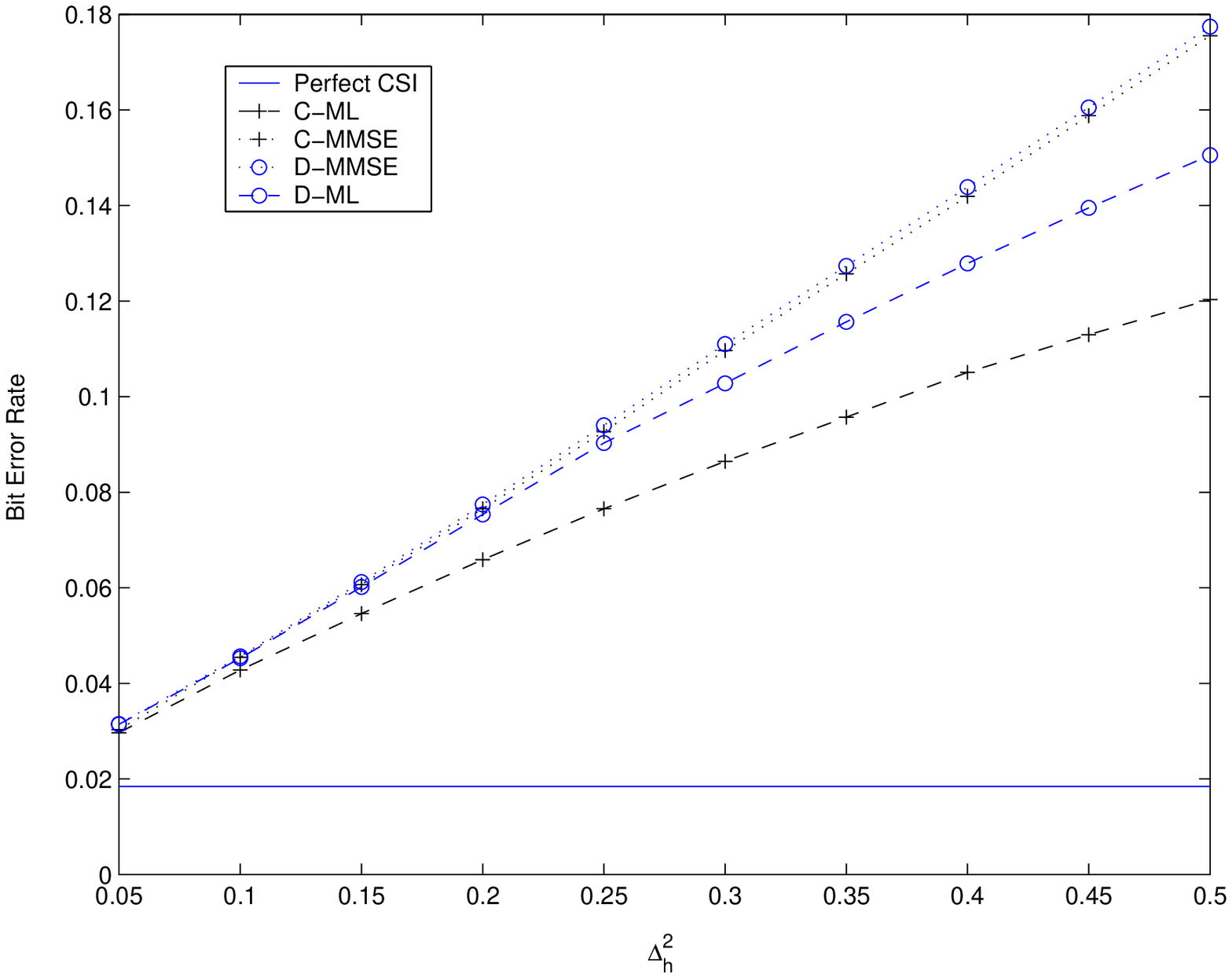}
  \caption{Bit error rate of C-IO MUD as a function of channel estimation error variance}\label{}
\end{figure}

\begin{figure}
  \centering
  \includegraphics[scale=1]{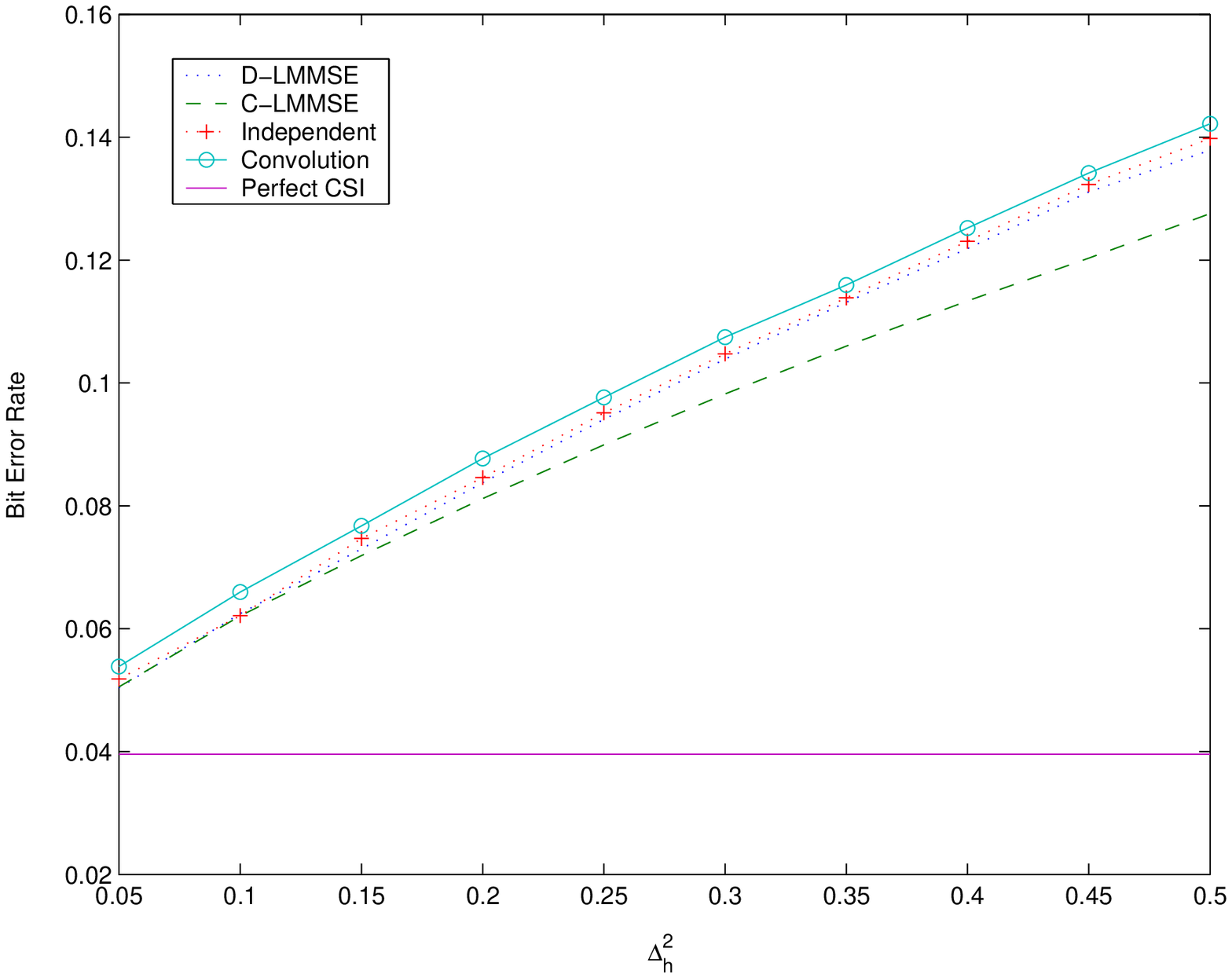}
  \caption{Bit error rate of MMSE MUD as a function of channel estimation error variance}\label{}
\end{figure}

\begin{figure}
  \centering
  \includegraphics[scale=1]{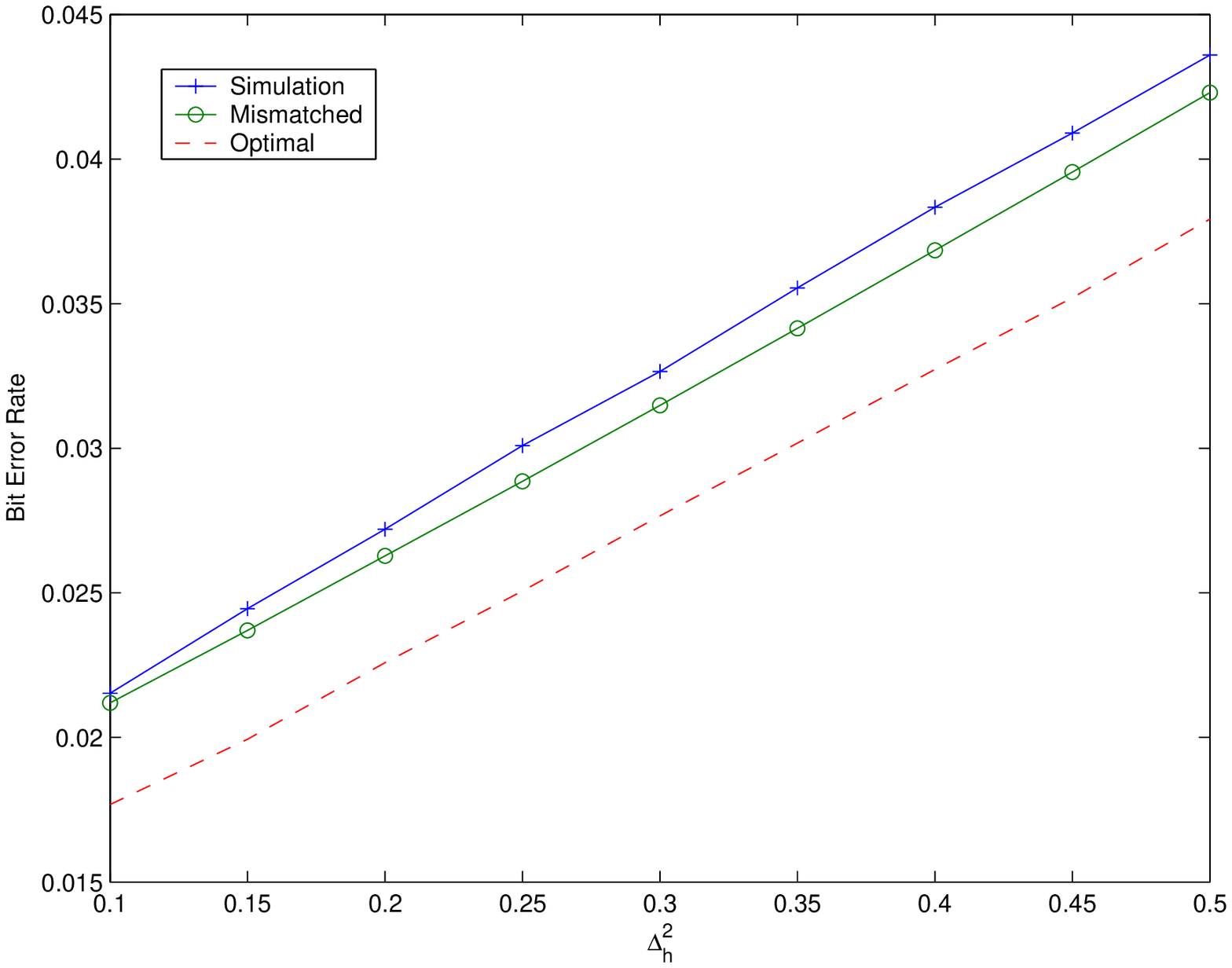}
  \caption{Bit error rate of MMSE filter based PIC as a function of channel estimation error variance}\label{}
\end{figure}

\begin{figure}
  \centering
  \includegraphics[scale=1]{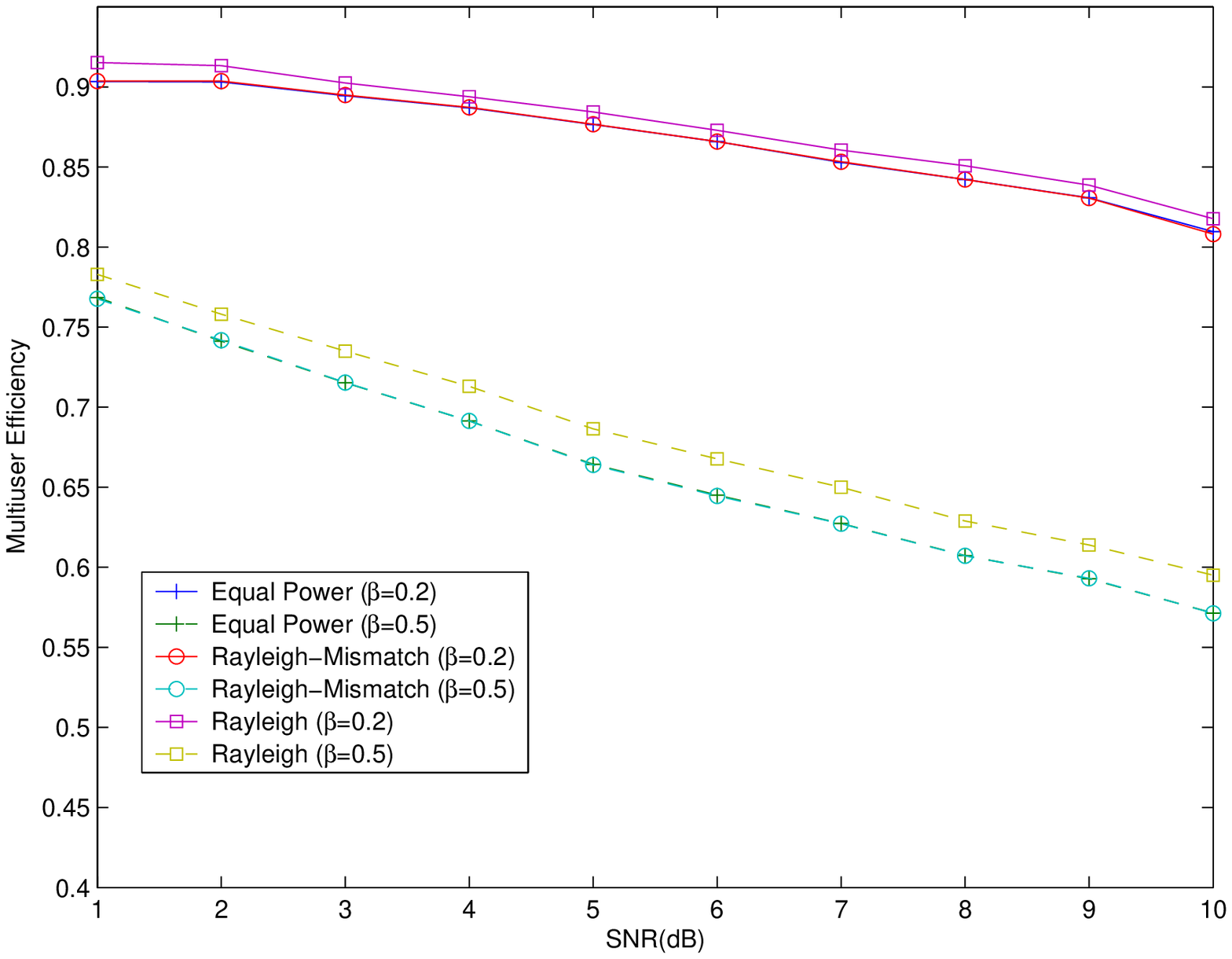}
  \caption{Multiuser efficiency versus SNR for non-fading and Rayleigh fading systems}\label{}
\end{figure}

\begin{figure}
  \centering
  \includegraphics[scale=1]{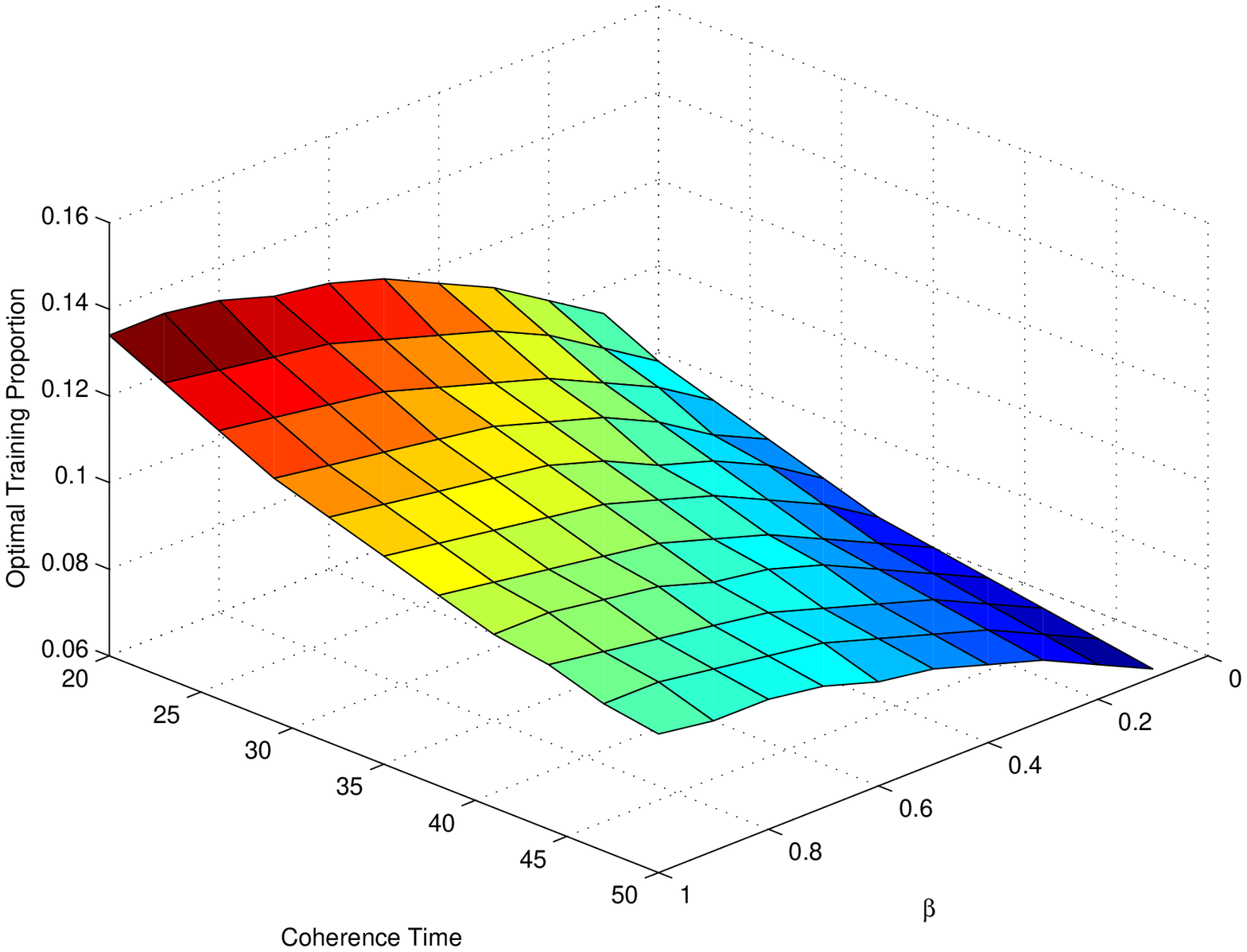}
  \caption{Optimal proportion of training symbols versus coherence times and system load}\label{}
\end{figure}


\begin{thebibliography}{11}
\bibitem{Alex2000}
P.~Alexander and A.~Grant, ``Iterative detection in code-division
multiple-acess with error control coding,'' {\em European Trans.
Telecommun.}, Vol.~9,
  pp.~419-426, Aug. 1998.

\bibitem{Caire2003}
G. Caire, R. M\"{u}ller and T. Tanaka, ``Iterative multiuser joint
decoding: Optimal power allocation and low-complexity
implementation,'' {\em IEEE Trans. Inform. Theory}, Vol.~50,
  pp.~1950-1973, Sept. 2004.

\bibitem{Evans2000}
J. Evans and D. N. C. Tse, ``Large system performance of linear
multiuser receivers in multipath fading channels,'' {\em IEEE
Trans. Inform. Theory}, Vol.~46,
  pp.~2059-2078, Aug. 2000.

\bibitem{GuoCISS2002}
D. Guo and S. Verd\'{u}, ``Multiuser detection and statistical
mechanics,'' in {\em Communications, Information and Network
Security} (V. Bhargava, H. V. Poor, V. Tarokh, and S. Yoon, eds.),
ch. 13, pp. 229-277, Kluwer Academic Publishers, Norwell, MA,
2002.

\bibitem{GuoCISS2003}
D.~Guo and S.~Verd\'{u}, ``Spectral efficiency of large-system
CDMA via statistical physics,'' {\em Proc. 2003 Conference on
Information Sciences and Systems}, The Johns Hopkins University,
Baltimore, MD, March 12-14, 2003.

\bibitem{Hollander2000}
F. D. Hollander, {\em Large Deviations}.
\newblock American Mathematical Society, Providence, RI, 2000.

\bibitem{LiHuCISS2003}
H. ~Li and H. V. Poor, ``Performance of channel estimation in long
code DS-CDMA with and without decision feedback,'' {\em Proc. 2003
Conference on Information Sciences and Systems}, The Johns Hopkins
University, Baltimore, MD, March 12-14, 2003.

\bibitem{LiHuWCNC2004}
H. Li and H.~V. Poor, ``Impact of imperfect channel estimation on
turbo MUD in DS-CDMA systems,''
 {\em Proc. 2004 IEEE Wireless Communications and
Networking Conference}, Atlanta, GA, March 21 - 25, 2004.

\bibitem{Lupas1989}
R.~Lupas and S.~Verd\'{u}, ``Linear multiuser detectors for
synthronous code-division multiple-acess channels,'' {\em IEEE
Trans. Inform. Theory}, Vol.~35,
  pp.~123--136, Aug. 1989.


\bibitem{Nishimori2001}
H. Nishimori, {\em Statistical Physics of Spin Glasses and
Information Processing}.
\newblock Oxford University Press, Oxford, UK, 2001.

\bibitem{Poor1998}
H. V. Poor, {\em An Introduction to Signal Detection and
Estimation}.
\newblock Springer-Verlag Press, New York, NY, 1994.


\bibitem{Silverstein}
K. W. Silverstein, ``Strong convergence of the empirical
distribution of eigenvalue of large dimensional random matrices,''
 {\em J. Multivariate Annl.}, vol. 55, no. 2, pp. 331-339, 1995.

\bibitem{Tanaka2002}
T. Tanaka, ``A statistical-mechanics approach to large-system
analysis of CDMA multiuser detectors,'' {\em IEEE Trans. Inform.
Theory}, Vol.~48,
  pp.~2888--2910, Nov. 2002.

\bibitem{Tse1999}
D.~Tse and S.~Hanly, ``Linear multiuser receivers: Effective
interference, effective bandwidth and user capacity,'' {\em IEEE
Trans. Inform. Theory}, Vol.~45,
  pp.~641-657, March 1999.

\bibitem{Verdu1984}
S. Verd\'{u}, {\em Optimal Multi-user Singal Detection.} PhD
thesis, University of Illinois at Urbana-Champaign, Aug. 1984.

\bibitem{Verdu1986}
S. Verd\'{u}, ``Minimum probability of error for asynchronous
Gaussian multiple-access channels,'' {\em IEEE Trans. Inform.
Theory}, Vol.~32,
  pp.~85--96, Jan. 1986.

\bibitem{Verdu1998}
S. Verd\'{u}, {\em Multiuser Detection}.
\newblock Cambridge University Press, Cambridge, UK, 1998.

\bibitem{Verdu1999}
S. Verd\'{u} and S. Shamai, ``Spectral efficiency of CDMA with
random spreading,'' {\em IEEE Trans. Inform. Theory}, Vol.~45,
  pp.~622--640, March 1999.

\bibitem{Voiculescu1992}
D. V. Voiculescu, K. J. Dykema and A. Nica, {\em Free Random
Variables}.
\newblock CRM Monograph Series, Volume 1, Providence, Rhode Island, USA: American Mathematical Society, 1992.

\bibitem{wang1999}
X. Wang and H. V. Poor, ``Iterative (turbo) soft interference
cancellation and decoding for coded CDMA,'' {\em IEEE Trans.
Commun.}, Vol.~47,
  pp.~1046-1061, Aug. 1999.

\bibitem{ZhengyuanXu2004}
Z. Xu, ``Effects of imperfect blind channel estimation on
performance of linear CDMA receivers,'' {\em IEEE Trans. Signal
Processing}, Vol.~52,
  pp.~2873-2884, Oct. 2004.
\end{thebibliography}
\end{document}